\documentstyle[epsf]{elsart}

\def\l{\langle}
\def\r{\rangle}

\begin{document}
\begin{frontmatter}
\title{
Cluster Analysis of the Ising Model \\
and Universal Finite-Size Scaling
}

\author[Tokyo]{Yutaka Okabe\thanksref{okabe}},
\author[Tokyo]{Kazuhisa Kaneda},
\author[Tokyo]{Yusuke Tomita},
\author[Osaka]{Macoto Kikuchi},
\author[Taipei]{Chin-Kun Hu}

\address[Tokyo]{
Department of Physics, Tokyo Metropolitan University,
Hachioji, Tokyo 192-0397, Japan
}

\address[Osaka]{
Department of Physics, Osaka University, Osaka 560-0043, Japan
}

\address[Taipei]{
Institute of Physics, Academia Sinica, Nankang, Taipei 11529, Taiwan
}

\thanks[okabe] {Electronic address: okabe@phys.metro-u.ac.jp}

\date{Received 15 October 1999}

\begin{abstract}
The recent progress in the study of finite-size scaling (FSS) 
properties of the Ising model is briefly reviewed.  
We calculate the universal FSS functions for the Binder 
parameter $g$ and the magnetization distribution function $p(m)$
for the Ising model on $L_1 \times L_2$ two-dimensional 
lattices with tilted boundary conditions.
We show that the FSS functions are universal 
for fixed sets of the aspect ratio $L_1/L_2$ and 
the tilt parameter.  
We also study the percolating properties of the Ising model, 
giving attention to the effects of the aspect ratio of 
finite systems.  We elucidate the origin of the complex 
structure of $p(m)$ for the system with large aspect ratio 
by the multiple-percolating-cluster argument.
\end{abstract}


\end{frontmatter}

\section{Introduction}
In the study of critical phenomena, universality and scaling are 
two important concepts \cite{stanley71}. 
Finite-size scaling (FSS) has been increasingly important 
\cite{fisher70,cardy88}, due to the progress in the theoretical 
understanding of finite-size effects, and the application of 
FSS in the analysis of simulational results.
Privman and Fisher \cite{pf84} first proposed the idea of universal 
FSS functions with nonuniversal metric factors in 1984, 
but only recently, the universality of FSS functions 
has received much attention 
\cite{hlc95a,hl96,ok96,wh97,lhc98,lh98,hu98,hcl98,ok99,toh99}. 

The first strong support for the idea of universal FSS functions 
was reported for critical phenomena in geometric percolation 
models \cite{hlc95a}.  
Hu, Lin and Chen \cite{hlc95a} applied a histogram Monte Carlo 
simulation method \cite{hu92b} to calculate the existence probability 
$E_p$ and the percolation probability $P$ of site and bond percolation 
on finite square (sq), plane triangular (pt), and honeycomb (hc) 
lattices, whose aspect ratios approximately have the relative proportions 
$1:\sqrt{3}/2:\sqrt{3}$ \cite{lpps92}.  They found that the six 
percolation models have very nice universal FSS for $E_p$ and $P$
near the critical points of the percolation models with 
using nonuniversal metric factors.
Using Monte Carlo simulation, Okabe and Kikuchi \cite{ok96} found 
universal FSS for the Binder parameter $g$ \cite{binder81} 
and magnetization distribution functions $p(m)$ of the Ising model 
on planar lattices; Wang and Hu discussed universal FSS for 
dynamical critical phenomena of the Ising model \cite{wh97}.

It is to be noted that universal FSS functions 
depend on boundary conditions and the shape of finite systems.
The importance of the number of percolating clusters 
for anisotropic systems was pointed out 
by Hu and Lin \cite{hl96,lh98}.  
The probability for the appearance of $n$ percolating clusters $W_n$ 
was investigated; the average number of percolating clusters 
increases linearly with aspect ratios of the lattices at the critical 
point.  Pioneered by the work of Hu and Lin \cite{hl96,lh98}, 
the number of percolating clusters has captured current interest 
\cite{aizenman97,ha97,cardy98}.  
Moreover, the ``nonuniversal scaling'' of the low-temperature 
conductance peak heights for Corbino disks in the quantum Hall effect 
was discussed in terms of the number of 
the percolating clusters \cite{chhr97}.
As for the effect of boundary conditions, 
the difference of the universal FSS functions between the systems 
with periodic boundary conditions and those with free boundary 
conditions was shown both in the percolation problem \cite{hlc95a} 
and the Ising model \cite{ok96}.

Another interesting subject is the relation of the geometric 
percolation problem to the phase transition problem of 
the spin models.  The understanding of critical phenomena 
of spin models in terms of geometric concepts has been enhanced 
by the cluster formalism introduced 
by Kasteleyn and Fortuin \cite{kf69}.  The problem of 
the thermal phase transition is mapped onto the geometric 
percolation problem in the cluster formalism 
\cite{kf69,ck80,hu84}.  The connection between the 
bond-correlated percolation model (BCPM) and the $q$-state Potts model 
was elucidated by Hu \cite{hu84}; 
an equation for the spontaneous magnetization was 
formally derived by including the magnetic field in the subgraph
expansion for the partition function of the $q$-state Potts model. 
The cluster formalism has been applied to the cluster update 
algorithms \cite{sw87,wolff88} in order to overcome the 
slow dynamics in the Monte Carlo simulation. 
It is important to study multiple percolating clusters 
for the spin models of anisotropic systems; 
especially, the aspect ratio dependence of the magnetization 
distribution function of the spin system is interesting to 
discuss in terms of the cluster formalism. 

In the first part of this paper, we study the universal FSS functions 
of the Ising model with tilted boundary conditions.  
We find that the FSS functions are universal for fixed sets of 
aspect ratio $a$ and tilt parameter $c$. 
This part of work has been partially reported elsewhere \cite{ok99}. 
We also discuss the relevance to the modular transformation 
\cite{cardy86}. 
In the second part we study the percolating property of the Ising 
model based on the connection between the BCPM and the Ising model.
Special attention is paid to the number of percolating 
clusters and its dependence on the aspect ratio of the lattice.  
We discuss the origin of the complex structure of the 
magnetization distribution function $p(m)$ 
for the system with large aspect ratio.  
The details of this part have been given 
in a separate paper \cite{toh99}. 

\section{Ising model with tilted boundary conditions}

\begin{figure}
\centerline{\epsfxsize=8.0cm \epsfbox{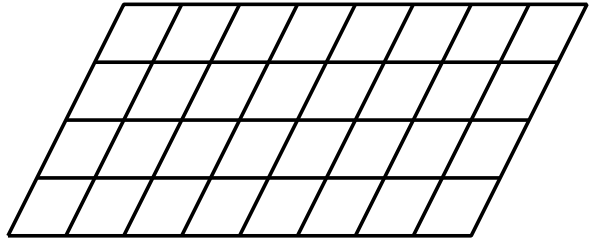}}
\vspace{2mm}
\caption{$L_1 \times L_2$ square lattice with tilt
 parameter $c$. Here, $L_1=8$, $L_2=4$ and $c=1/4$.  
 The $i$-th site of the first row is identical with 
 the mod$(i + c L_1,L_1)$-th site in the last row.
 The left-most site and the right-most site 
 on the same horizontal line are identical.}
\label{tilt}
\end{figure}

We deal with the two-dimensional (2D) Ising model 
on $L_1 \times L_2$ sq lattices 
with periodic boundary conditions in the horizontal $L_1$ 
direction and tilted boundary conditions in the vertical $L_2$ 
direction such that the $i$-th site in the first row is connected 
with the mod$(i + c L_1,L_1)$-th site in the $L_2$ row of the lattice, 
where $1 \le i \le L_1$; see Fig.~\ref{tilt} for an example.
We mainly calculate the universal FSS functions for 
the Binder parameter \cite{binder81}
\begin{equation}
 g=\frac{1}{2}(3-\frac{\l m^4 \r}{\l m^2 \r^2})
\label{binder_ratio}
\end{equation}
and the magnetization distribution function $p(m)$, 
using the Monte Carlo simulation method. 

\begin{figure}
\centerline{\epsfxsize=10.0cm \epsfbox{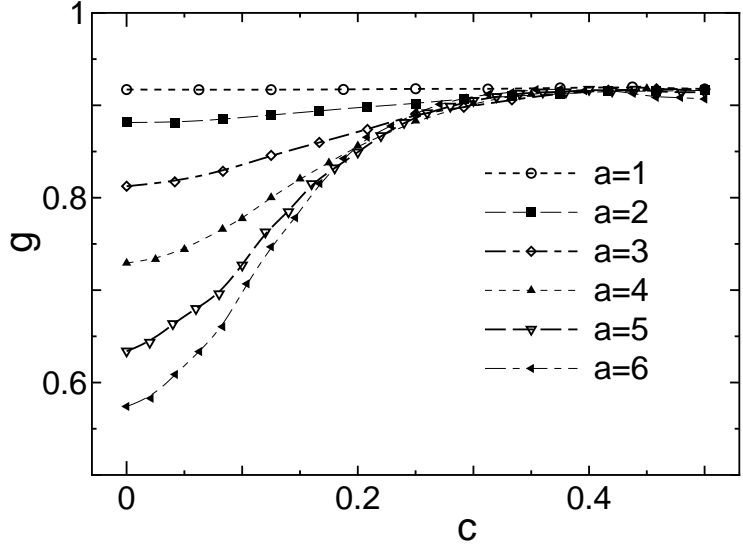}}
\vspace{2mm}
\caption{$g$ at $T=T_c$ as a function of $c$ 
for several values of $a=L_1/L_2$.}
\label{g_depend}
\end{figure}

We have found very good FSS behavior for $g$ as a function of 
$(T-T_c) L^{1/\nu}$ and $p(m)L^{-\beta/\nu}$ as a function of 
$m L^{\beta/\nu}$ \cite{ok99}. 
Here, $L=(L_1 L_2)^{1/2}$, $T_c = 2.269 \cdots$ in units of 
the exchange coupling $J$, and 
the critical exponents are 
those of the 2D Ising exact values; $\nu=1$ and $\beta=1/8$.
Moreover, FSS functions of $g$ and $p(m)$ have been shown to 
depend strongly on the tilt parameter $c$.  
In order to see the tilt parameter dependence clearly, 
we plot the $c$ dependence of $g$ at $T=T_c$ for several values 
of the aspect ratio $a=L_1/L_2$ in Fig.~\ref{g_depend}.

\begin{figure}
\centerline{\epsfxsize=10.0cm \epsfbox{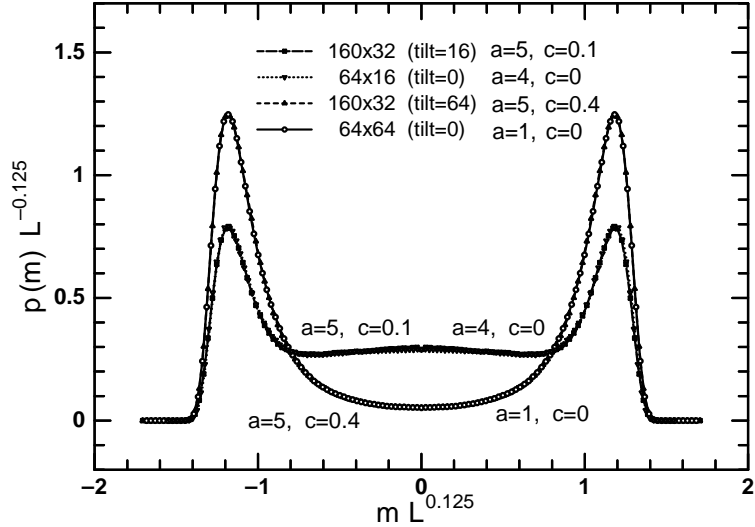}}
\vspace{2mm}
\caption{$p(m) L^{-1/8}$ at $T=T_c$ as a function of $mL^{1/8}$ for 
$(a,c)$ = (5,0.1), (4,0), (5,0.4), and (1,0).}
\label{universal}
\end{figure}

We find a combination of $a$ and $c$ which gives 
universal FSS functions.  As an example, we show the data 
for $p(m)$ at $T=T_c$ for $(a,c)$ of (5,0.1), (4,0), (5,0.4), 
and (1,0) together in Fig.~\ref{universal}.  
The figure shows that the pair (5,0.1) and (4,0) and the pair
(5,0.4) and (1,0) share the universal FSS functions.  
There are many such combinations that share universal FSS functions.

Using the FSS argument in the momentum space, 
we have shown \cite{ok99} that 
\begin{equation}
A=a/(c^2 a^2+1)
\label{invariant}
\end{equation}
is an invariant, and can be regarded as 
the effective aspect ratio.  
It is easy to check that the pairs of $(a,c)$, 
which have the universal FSS functions shown 
in Fig.~\ref{universal}, have the same value of $a/(c^2 a^2+1)$.
The invariant, Eq.~(\ref{invariant}), 
corresponds to the relative ratio of two primitive vectors 
${\bf k}_1$ and ${\bf k}_2$ in the momentum space. 
More precisely, we may find other expressions than Eq.~(\ref{invariant}) 
for the invariant in case of some values of $a$ and $c$; 
the details will be given elsewhere. 

It is interesting to discuss this problem in terms of the 
modular (conformal) transformation.  According to Cardy \cite{cardy86}, 
the shape of the 2D lattice may be represented by the imaginary number
\begin{equation}
  z = 1/a + i \ c.
\label{imaginary}
\end{equation}
Then, Cardy asserted that the partition function becomes invariant 
under the transformations
\begin{equation}
  z \rightarrow z + i
\end{equation}
and
\begin{equation}
  z \rightarrow 1/z,
\end{equation}
in the limit that the system size becomes infinite.
We have another invariant transformation
\begin{equation}
  z \rightarrow z^*,
\end{equation}
which corresponds to the fact that we can confine $c$ to the interval 
of $0 \le c \le 1/2$.
Starting from the recurrence relation 
\begin{equation}
  z_{n+1} = \frac{1}{z_n+i} + i,
\end{equation}
we can easily show that Eq.~(\ref{invariant}) becomes invariant. 
What we found in the Monte Carlo simulation is that the FSS 
functions are universal under the modular transformations 
based on the imaginary-number representation, Eq.~(\ref{imaginary}). 
The detailed analysis of the modular transformation is now 
in progress.

We make a comment here that Ziff, Lorentz and Kleban \cite{zlk99} 
have recently studied the universal excess cluster numbers 
for percolation on lattices with tilted (twisted) boundary conditions.
The related mathematical problems are also of interest to mathematical 
physicists.

\section{Cluster analysis of the Ising model}
Here, we study the FSS properties of the percolating clusters 
for the 2D Ising model based on 
the connection between the BCPM and the Ising model.  
The partition function of the Ising model, 
generally that of the $q$-state Potts model, can be expressed by 
the BCPM \cite{kf69,ck80,hu84}. 
The essential point is that the problem of the Ising model is 
mapped to the percolation problem with the bond concentration of 
\begin{equation}
     p=1-e^{-2J/T}, 
\label{p}
\end{equation}
where $J$ is the nearest-neighbor exchange coupling.  

We perform the Monte Carlo simulation of the Ising model 
on the $L_1 \times L_2$ sq lattice with the periodic boundary conditions. 
We denote the aspect ratio as $a=L_1/L_2$ again.  For the 
assignment of a bond-percolating cluster, we consider free and periodic 
boundary conditions in the vertical and horizontal directions, 
respectively; that is, a cluster which extends from the top row 
to the bottom row is a percolating cluster.

\begin{figure}
\centerline{\epsfxsize=10.0cm \epsfbox{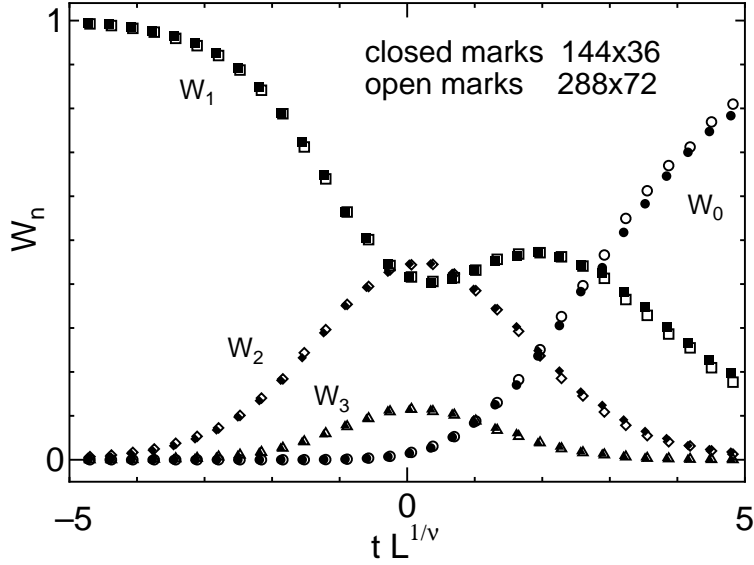}}
\vspace{2mm}
\caption{FSS plot of $W_n$ ($n$ = 0, 1, 2, 3) 
for the sq lattice of $144 \times 36$ (closed marks) and 
$288 \times 72$ (open marks), where $t=(T-T_c)/T_c$.
}
\label{Wn}
\end{figure}

We calculate the probability for the appearance of $n$ 
percolating clusters $W_n$ for anisotropic lattices \cite{hl96,lh98}. 
This quantity is related to the existence probability $E_p$ as
\begin{equation}
  \sum_{n=1}^{\infty}W_n = 1-W_0 = E_p. 
\end{equation}
In Fig.~\ref{Wn}, we show the FSS plot of $W_n$ as a function of 
$tL^{1/\nu}$, where $t=(T-T_c)/T_c$. 
The closed and open marks are the data for $144 \times 36$ and for 
$288 \times 72$, respectively.  The aspect ratio $a$ is 4. 
We see interesting behavior for $W_1$; that is, from the high-temperature 
side $W_1$ increases as the temperature approaches the critical
temperature, but decreases near $T_c$.  
Moreover, $W_1$ increases again for low temperatures, and finally 
approaches 1.  This is the same behavior observed for the
geometric percolation problem \cite{hl96,lh98}.  

\begin{figure}
\centerline{\epsfxsize=10.0cm \epsfbox{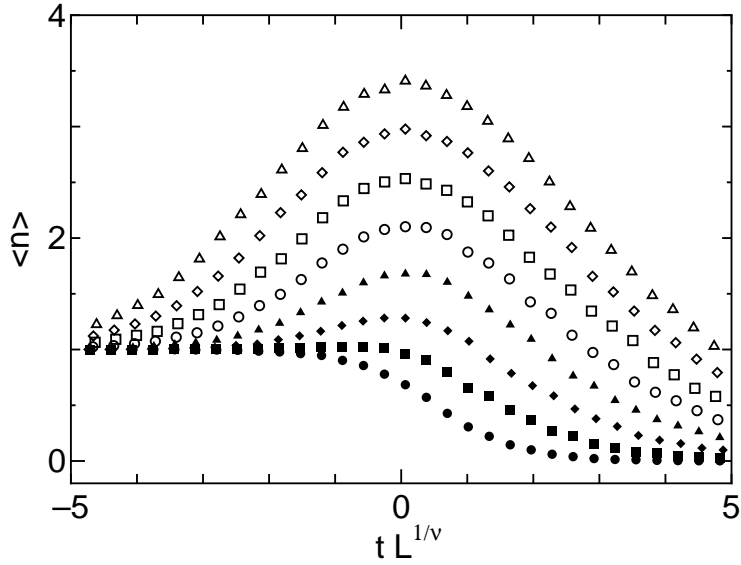}}
\vspace{2mm}
\caption{Temperature dependence of $\l n \r$ 
for various aspect ratios $a = L_1/L_2$; 
$a = 8, 7, 6, 5, 4, 3, 2, 1$ from top to bottom.}
\label{n_av}
\end{figure}

The average value $\l n \r=\sum_n nW_n$ gives the direct measure 
of number of percolating clusters.  
To investigate the aspect ratio dependence, 
we give the temperature dependence of $\l n \r$ 
for various aspect ratios $a=L_1/L_2$ from 1 to 8 in Fig.~\ref{n_av}.  
For high temperatures, $\l n \r$ is small because there is no order.  
Then, $\l n \r$ takes a maximum value near $T_c$, and 
decreases to 1 as $T$ lowers.  There is only 
a single percolating cluster at low enough temperature.
A systematic $a$ dependence is observed for the behavior 
of $\l n \r$.

\begin{figure}
\centerline{\epsfxsize=9.0cm \epsfbox{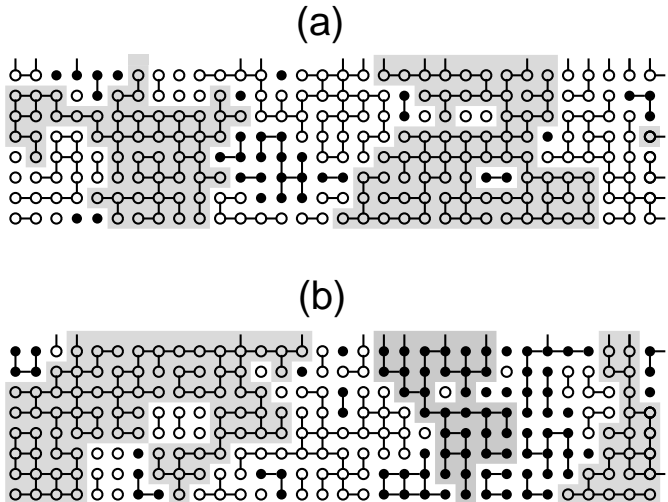}}
\vspace{2mm}
\caption{
Examples of snapshots of the Ising system with the aspect ratio 
$a=4$.  Up and down spins are represented by open and closed circles, 
and active bonds are represented by solid line. 
Percolating clusters are distinguished by shaded area.  
There are two percolating clusters with up spins in (a), whereas 
one percolating cluster is up and the other is down in (b).}
\label{snap}
\end{figure}

It is known that the magnetization distribution function $p(m)$ 
depends on the shape of the finite systems.  In Fig.~\ref{universal}, 
we showed that the FSS functions for $a=4$ and $a=1$ are quite different.  
For the system with the aspect ratio $a=4$, 
the total distribution function $p(m)$ at $T=T_c$ 
has a broad peak centered at $m=0$ in addition to two peaks of plus 
and minus $m$;  this is contrast to the case of $a=1$ where $p(m)$ 
has only two distinct peaks of plus and minus $m$.  
It is quite interesting to study this dependence on 
the aspect ratio by means of the cluster analysis.
We may decompose $p(m)$ by the number of 
percolating clusters as 
\begin{equation}
  p(m) = \sum_{n=0}^{\infty}p_n(m).  
\end{equation}
It should be noted that we have the relation, 
\begin{equation}
  \int_{-1}^1 p_n(m)dm = W_n, \quad (n=0, \cdots, \infty),  
\end{equation}
to connect $p_n(m)$ and $W_n$.
By analyzing the FSS plot of $p_n(m)$ at $T=T_c$, 
we found that a broad peak centered at $m=0$ mainly comes from $p_2(m)$
for $a=4$ \cite{toh99}.  There are two types of Ising clusters, that is, 
the clusters with up spins or the clusters with down spins.  
Therefore, if there are many percolating clusters, the combination of 
the percolating clusters with up spins and those with down spins 
gives the contribution to the broad peak around $m=0$ in $p(m)$. 
Examples of snapshots of the Ising system with two percolating clusters 
are presented in Fig.~\ref{snap}. 
In case of (a), both percolating clusters are up;  
in contrast, in case of (b), one percolation cluster is up 
and another percolating cluster is down. 
It is known that the Binder parameter, Eq.~(\ref{binder_ratio}), 
at the critical point has an aspect-ratio 
dependence \cite{ka93}.  The origin of such a dependence 
can be attributed to the structure of multiple percolating clusters.

In order to check the universal FSS, we have also made the 
calculation for the Ising model on the pt and hc lattices, 
whose aspect ratios are approximately $4 \times \sqrt{3}/2$ 
and $4 \times \sqrt{3}$, respectively.  
We have obtained very good universal FSS for the 
temperature dependence of $W_n$ and other quantities.
The details of the calculation have been reported in 
Ref.~\cite{toh99}.

\section{Summary and discussions}

We have studied the universal FSS functions for the Ising model 
paying attention to the effects of the boundary conditions and 
the aspect ratio.  For the system with the tilted boundary conditions 
we have found a combination of the aspect ratio $a$ and the tilt 
parameter $c$ which gives universal FSS functions.  
We have discussed the relevance to the modular transformation 
in 2D systems \cite{cardy86}.

We have also studied the FSS for the percolating properties of 
the Ising model based on the connection between the BCPM and the 
Ising model.  Investigating the multiple percolating clusters 
for the system with large aspect ratio, we have clarified 
the origin of the complex structure of $p(m)$ near $T_c$.  
This is due to the combination of the clusters with up spins 
and those with down spins.  

The extention of the present study to higher-dimensional systems is 
quite important. 
Conformal invariance plays a role in 2D systems \cite{cardy87}, 
but is not so powerful for three-dimensional (3D) systems as for 2D ones.  
We have studied the FSS functions for anisotropic 
3D Ising models of size $L_1 \times L_1 \times aL_1$ 
by Monte Carlo simulations \cite{kok99}.  
We have observed the change of the FSS functions for $p(m)$ at $T_c$ 
as a function of $a$, which is the same situation as 2D systems.

Another interesting problem is the application of the present study 
to random spin systems.  
The study of the percolating properties of the diluted Ising model, 
especially the crossover from the percolation fixed point 
to the Ising fixed point, is now in progress.

\section*{Acknowledgments}
We would like to thank T. Kawakatsu, N. Hatano and 
N. Kawashima for valuable discussions, and 
the Supercomputer Center of the ISSP, University of 
Tokyo, for providing the computing facilities.
This work was supported by a Grant-in-Aid for 
Scientific Research from the Ministry of Education, 
Science, Sports and Culture, Japan and by the National 
Science Council of the Republic of China (Taiwan) 
under grant numbers NSC 88-2112-M-001-011.

\end{document}